\documentclass[reprint,5p]{elsarticle}
\usepackage[usenames,dvipsnames]{color}
\usepackage[normalem]{ulem}
\usepackage{amsmath}
\usepackage{todonotes}
\usepackage[capitalize]{cleveref}
\usepackage[acronym]{glossaries}
\usepackage{listings}
\usepackage{url}
\usepackage{lineno}
\usepackage{microtype}
\usepackage{xcolor}
\usepackage[binary-units]{siunitx}

\newcommand\hl{\bgroup\markoverwith
  {\textcolor{yellow}{\rule[-.5ex]{2pt}{2.5ex}}}\ULon}

\def\backtick{\char18}
\lstdefinestyle{mystyle}{literate={`}{\backtick}1, escapechar=@}
    
\usepackage{booktabs}
\usepackage{tabularx}

\makeatletter
\providecommand{\doi}[1]{doi:#1}
\makeatother
\biboptions{compress}

\makeglossaries

\title{Simple Data and Workflow Management with the \textbf{signac} Framework}
\author[umichche]{Carl~S.~Adorf}
\ead{csadorf@umich.edu}

\author[umichche]{Paul~M.~Dodd}
\ead{pdodd@umich.edu}

\author[umichche]{Vyas~Ramasubramani}
\ead{vramasub@umich.edu}

\author[umichche,umichmse,umichbi]{Sharon~C.~Glotzer\corref{cor1}}
\ead{sglotzer@umich.edu}

\cortext[cor1]{Corresponding author}

\address[umichche]{Department of Chemical Engineering, University of Michigan, Ann Arbor, MI 48109}
\address[umichmse]{Department of Materials Science and Engineering, University of Michigan, Ann Arbor, MI 48109}
\address[umichbi]{Biointerfaces Institute, University of Michigan, Ann Arbor, MI 48109}

\newcommand{\signac}{\texttt{signac}}
\newcommand{\sflow}{\texttt{signac-flow}}
\newcommand{\sdashboard}{\texttt{signac-dashboard}}

\newcommand{\package}[1]{\texttt{#1}}

\newcommand{\fp}{FlowProject}
\newcommand{\fps}{\fp{}s}
\newcommand{\hoomd}{\texttt{HOOMD-blue}}
\newcommand{\mongodb}{MongoDB}

\newacronym{api}{API}{Application Program Interface}
\newacronym{cli}{CLI}{command-line interface}
\newacronym{json}{JSON}{JavaScript Object Notation}
\newacronym{lj}{LJ}{Lennard-Jones}
\newacronym{gui}{GUI}{graphical user interface}
\newacronym{sql}{SQL}{Structured Query Language}
\newacronym{pdb}{PDB}{Protein Data Bank}
\newacronym{csd}{CSD}{Cambridge Structural Database}
\newacronym{hpc}{HPC}{high-performance computing}
\newacronym{dbms}{DBMS}{database management systems}
\newacronym{irods}{iRODS}{Integrated Rule-Oriented Data System}
\newacronym{nosql}{NoSQL}{non relational}
\newacronym{i/o}{I/O}{input/output}
\newacronym{srt}{SRT}{system response time}
\newacronym{nist}{NIST}{National Institute of Standards and Technology}
\newacronym{chimad}{CHiMaD}{Center for Hierarchical Materials Design}

\begin{document}

\begin{abstract}
Researchers in the field of materials science, chemistry, and computational physics are regularly posed with the challenge of managing large and heterogeneous data spaces.
The amount of data increases in lockstep with computational efficiency multiplied by the amount of available computational resources, which shifts the bottleneck in the scientific process from data acquisition to data processing and analysis.
We present a framework designed to aid in the integration of various specialized data formats, tools and workflows.
The \signac\ framework provides all basic components required to create a well-defined and thus collectively accessible and searchable data space, simplifying data access and modification through a homogeneous data interface that is largely agnostic to the data source, \textit{i.e.}, computation or experiment.
The framework's data model is designed to not require absolute commitment to the presented implementation, simplifying adaption into existing data sets and workflows.
This approach not only increases the efficiency with which scientific results can be produced, but also significantly lowers barriers for collaborations requiring shared data access.
\end{abstract}


    \maketitle


\printglossaries

\section{Introduction}
\label{sec:introduction}

\begin{figure}
    \centering
    \includegraphics[width=0.50\linewidth]{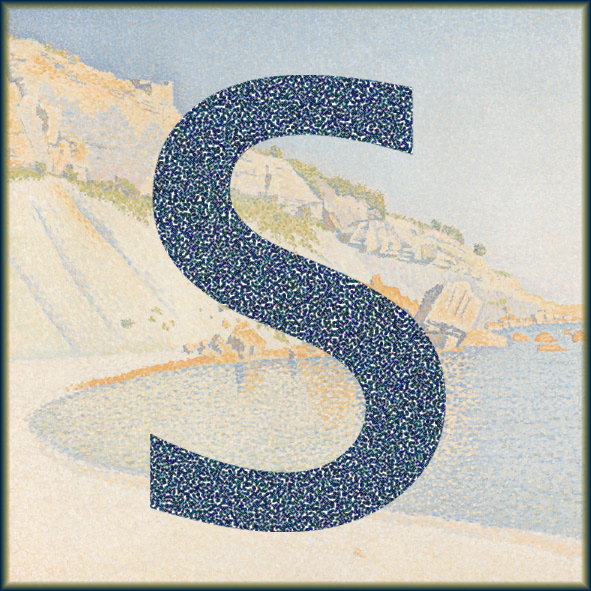}
    \caption{
        The Pointillist style was invented by Paul Signac (1863-1935) and Georges Seurat (1895-1891) and describes paintings in which images are composed from collections of individual dots, each containing a single color.
        This style serves as a metaphor for \signac's data model, in which the data is dependent on both individual data points \emph{and} their position within the larger parameter space.
        The painting underlying this artistic illustration \emph{Cassis, Cap Lombard} was created by Paul Signac in 1889 and is owned by the Gemeentenmuseum in Den Haag.
    }
    \label{fig:signac-logo}
\end{figure}

Improved software ~\cite{Plimpton1995,Anderson2013a,Anderson2013,Anderson2016,Abraham2015a} and increased resources available to computational researchers~\cite{Shirts2000,Towns2014} have led to significant increases in the quantities of data generated~\cite{Jain2013c}.
This makes a highly systematic data management approach crucial to preserving data provenance and ensuring reproducibility.
To address this problem, researchers often employ data organization practices such as using human-readable file-naming conventions.
Although such solutions address the problem at a superficial level, they suffer from numerous drawbacks with respect to efficiency and flexibility.
Here, we introduce \signac, named after Paul Signac (see \cref{fig:signac-logo}), a simple and robust framework for the management of complex and heterogeneous data spaces as well as the efficient implementation of workflows.
Data spaces managed with \signac\ are immediately searchable and sharable.

The capabilities of \signac\ are best illustrated by example.
Consider a typical, albeit trivial, research task in which we are given data about the pressure, volume, and temperature of a noble gas and wish to develop a simple theory to explain these data. 
As a first hypothesis, we might test Boyle's law, $pV = \text{const.}$, by iterating over values of $p$ and storing the corresponding values for $V$ in text files named for those values of $p$.
Upon finding that the data appears to be temperature-dependent, we then could choose to test a more general equation, $pV = NkT$.

We are now faced with a dilemma: how do we efficiently adapt our data space for this extension?
We could provide the existing files with new names incorporating temperature, but this could quickly become intractable if we had to further increase the complexity of our equation of state.
Alternatively, we might determine that storing data in a (relational) database would be a more flexible solution to accommodate any future schema changes; however, that could be much less efficient for a generally file-based workflow and could introduce a significant bottleneck in downstream data processing and analysis.

The \signac\ framework resolves this by abstracting away the details of file-based data storage while simultaneously functioning like a lightweight, semi-structured database.
Using \signac, files are directly stored on the file system \emph{along with the associated metadata} in a well-defined storage layout.
The metadata is parsed and indexed on-the-fly whenever we use \signac's interface to access and search for data.
By using \signac\ to manage the data in the above example, the tasks of adding a parameter such as temperature and searching for data associated with a particular ${p, T}$ pair can both be easily realized with only a few commands.

This paper is organized as follows.
First, the general design principles of \signac\ are presented.
We then delve into greater detail about how the core \signac\ functionality is implemented in keeping with these principles, followed by a more in-depth comparison to closely related solutions.
Finally, the practicality of this system is then demonstrated through numerous examples indicating how \signac\ can be used to manage a variety of disparate, heterogeneous data sets.

\section{Overview}

\subsection{Design}

In the following section we lay out the core design principles behind \signac, which necessitates making a clear distinction between the \signac\ \emph{framework} and the \signac\ \emph{application}.
The primary focus of this paper is the \signac\ application (henceforth simply \signac), which implements the core data management functions discussed throughout this paper.
The \signac\ framework is a collection of applications and modules that are built on top of the core \signac\ application, such as the \sflow\ application, which will be introduced in \cref{sec:signac-flow}.

\begin{figure}
    \centering
    \includegraphics[width=0.75\linewidth]{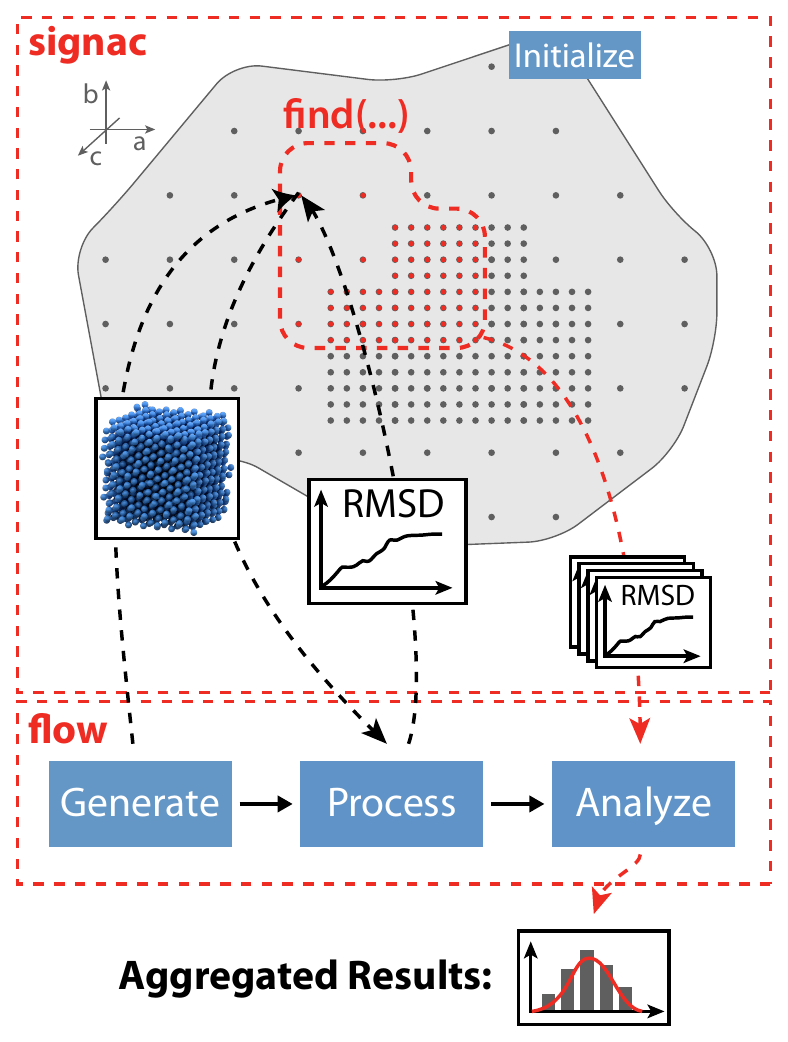}
    \caption{
        This conceptual example demonstrates how we manage and operate on a data space using the \signac\ and the \sflow\ applications.
        We use \signac\ to \emph{initialize} a discrete data space (represented by dark grey dots), where each dot represents a discrete data point and may be associated with anything from a single number to a large set of data.
        The data space is coordinated within a higher-dimensional parameter space (light grey shape), in this case spanned by the three vectors $a$, $b$, and $c$.
        Manipulations of the data space (addition, modification, or removal of data), can be divided into \emph{operations}, where each operation must be a function of one or more data points.
        The operations shown in the example deposit and extract data (dashed arrows) and are organized into a specific workflow using \sflow.
        Specifically, after initialization, we first \emph{generate} particle configurations, then \emph{post-process} these configurations to extract the root-mean squared displacement (RMSD).
        Finally, we aggregate results \textit{via} the \emph{analysis} of a subset of our data space that we \emph{find} using a \signac\ search query.
        This example shows the clear division of responsibilities between the different applications.
        The signac application manages and provides access to the data space and allows us to perform complex search queries.
        The \sflow\ application assists in the definition and execution of reproducible workflows comprised of individual data space operations.
        }
    \label{fig:dataspace}
\end{figure}

At its core, \signac\ is a database built directly on top of the file system, leveraging the many advantages of direct file system access while also providing functions to efficiently index and search the data space.
As a database system, \signac\ makes only one central assumption: that all data may be discretized within a high-dimensional parameter space (see \cref{fig:dataspace}).
Once the user provides the parameters and associated data, \signac\ is responsible for managing both the storage of data and its association with the parameters through the maintenance of metadata files encoded in the open \gls{json} format.
Through this division, \signac\ can ensure both data integrity and searchability.

The database functions of \signac\ are modeled after those provided by well-tried \gls{dbms} such as MongoDB~\cite{mongodb} or MySQL~\cite{mysql}.
Typically, such \gls{dbms} are very efficient when it comes to the execution of complex query and aggregation operations; however, there are two main issues that render these tools suboptimal for managing the large amounts of (binary) data typically generated by massively parallelized scientific applications within \gls{hpc}.
First, unless a database is specifically set up to handle peak loads originating from many instances (potentially numbering in the thousands) hitting them in parallel, reading and writing to files distributed on the file system will usually scale more efficiently.
Setting up a partitioned or replicated database system to handle higher loads is non-trivial, and this task becomes even more complicated if we care about proper authentication and authorization among different nodes.
Secondly, data may need to be serialized for ingestion into the database, which may pose another performance bottleneck, particularly if the data are large binary files.

With \signac, files are managed directly on the file system and performance is mainly determined by the latency and scalability of the file system.
This technique fully exploits the existing file systems on supercomputers, which are commonly designed to process highly parallel, computationally intensive \gls{i/o} operations, thereby avoiding all the above mentioned issues while also allowing for the immediate execution of the previously mentioned query routines.

\begin{figure}
    \centering
    \includegraphics[width=0.75\linewidth]{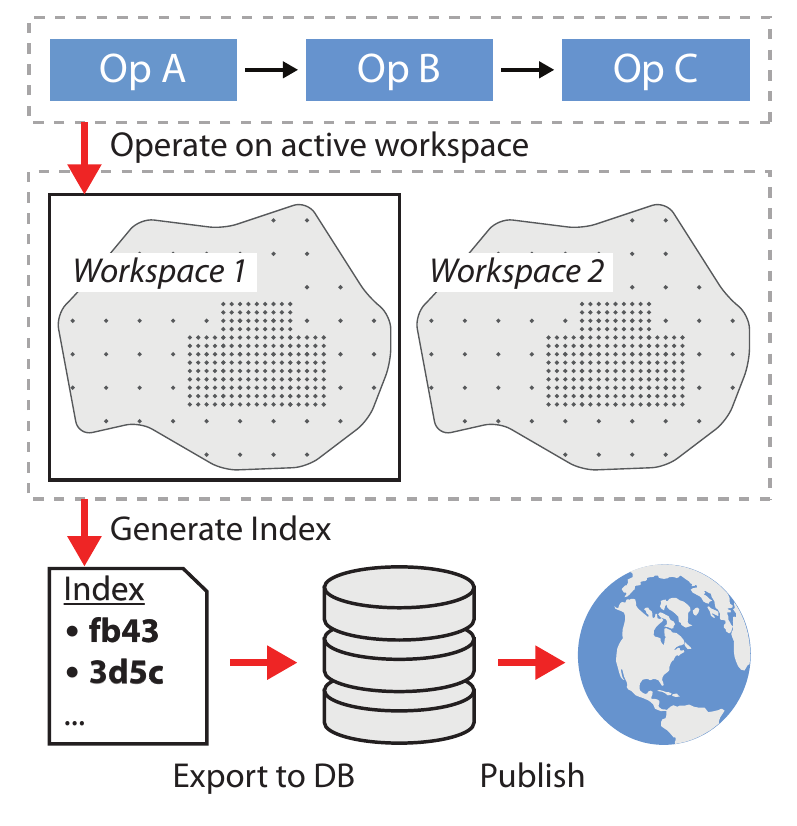}
    \caption{
        The \signac\ application manages a particular data space (illustrated in \cref{fig:dataspace}) by allocating it to a distinct \emph{workspace} (grey shaded space) on the file system.
        Data space operations (blue shaded boxes) used for the curation of data are always operating on one specific \emph{active} workspace (black frame).
        Information about state points, data location and data format may be compiled into an index using \signac.
        The index can be used for searching, aggregation, and even direct access to data.
        The index as well as the data itself, can be exported into a database, which is especially useful for the purpose of making data available to a wide range of subscribers, such as the general public.
        }
    \label{fig:workspaces-indexing}
\end{figure}

The \signac\ data model assumes that all data associated with a particular computational investigation is part of the same high-dimensional data space and therefore adheres to roughly the same semi-structured schema.
Each such investigation is called a \signac\ \emph{project}, and the associated data is stored in a special directory, the project \emph{workspace} (\cref{fig:workspaces-indexing}).
The data associated with any given set of parameters within the project's data space is sorted into a distinct subdirectory within the workspace along with a \gls{json} file containing the associated metadata.
In the introductory example, each ${p, T}$ pair represents a point within the larger parameter space, so the data associated with each pair would be stored in a distinct directory within the workspace along with a file containing the corresponding pressure and temperature.

This storage mechanism not only enables efficient on-the-fly indexing, it also ensures that parsing a \signac\ managed data space is straightforward even without \signac\ since the parameters associated with the data are stored at the same location.
In practice, however, \signac\ users can ignore these details since the software abstracts away the internal representation of the data space.
As described in \signac's public documentation\footnote{www.signac.io}, \signac\ enables users to easily access the high-level information required to interpret the data space without ever inspecting the filesystem directly.

None of this relieves the user of the burden of documenting their data spaces, \textit{i.e.}, describing explicitly the processes used to generate the data from the provided parameters; this procedure is facilitated by using \sflow.
Combined with proper documentation of these processes, however, the use of \signac\ ensures that a data space is fully interpretable even for individuals who did not create it.

This interpretability is critical because it makes the data accessible to anyone, even individuals not using \signac\ in their own workflows. 
There is strong evidence that well-maintained public databases, such as the \gls{pdb}~\cite{Berman2000}, the \gls{csd}~\cite{Allen2002,Groom2014}, The Materials Project~\cite{Jain2013c} or ImageNet~\cite{Deng2009} have a significant positive impact on their respective fields.
Promoting an open data culture among researchers within one or across multiple organizations will likely result in similar positive synergistic effects.
The simplicity of \signac\ is designed to facilitate this open data culture, because it lowers the barrier to adopt a standardized data storage layout, even for small data spaces and simple workflows that do not necessarily warrant a more sophisticated solution.
A data set managed with \signac\ that is uploaded to a repository such as the materials data facility\footnote{https://www.materialsdatafacility.org/} (the \gls{nist} and the \gls{chimad}) or the NOMAD repository\footnote{https://repository.nomad-coe.eu/} (funded by the European Union) is immediately easier to parse, access, and search.
A repository interface could be set up to directly support \signac, which would allow users to search the data by metadata directly.
Furthermore, any standardization of metadata tracking facilitates the curation and export of data to public databases such as the NRELMatDB\footnote{https://materials.nrel.gov/} or the materials data base\footnote{https://materialsdata.nist.gov/} managed by \gls{nist}, since converting an existing schema is easier than starting from scratch.

All of \signac's core functions are enabled through a highly efficient, \textit{on-the-fly} indexing of the data space.
For all higher-level functions, such as data searching and data selection, this indexing process is completely transparent to the user.
As a result, \signac\ maintains an extremely low barrier to entry, enabling new users to take immediate advantage of basic data management functions.
Meanwhile, more advanced users can access \signac's full range of capabilities (including detailed control over indexing) for the implementation of complex data-driven workflows.

To remain lightweight and focused, \signac\ does not attempt to solve \textit{all} data management concerns.
For example, we assume that infrastructure-related issues such as the setup of and access to a distributed file system are better addressed and solved by systems such as the \gls{irods}~\cite{irods} or GLOBUS~\cite{Foster2011}, both of which have a different scope than \signac.

\subsection{Workflow}
\label{sec:workflowmodel}

In order to support generic file-based workflows, the \signac\ data model makes minimal assumptions about how these workflows generate and operate on the data; \signac\ manages the file paths, but the underlying files are stored directly on the file system without modification or serialization.
This design ensures that existing tools may interact with a \signac\ data repository without the need to serialize or convert existing file formats, an advantage shared by solutions like datreant~\cite{datreant}.
Conversely, this design distinguishes \signac\ from more domain-specific solutions that make certain assumptions about data schema and format, such as DCMS~\cite{Kumar2014b} and the AiiDA infrastructure~\cite{Pizzi2016}.
See \cref{sec:survey} for a more detailed discussion.
Hence, a \signac\ workspace can be written to or read from outside the context of any broader workflow, and this framework can be used irrespective of how the data is generated or what must be done to process it as long as it is file-based.
In other words, whether data is generated through the evaluation of a single equation, or by means of compute-intensive molecular dynamics simulations, \signac\ is used in exactly the same way.

While \signac\ itself is workflow agnostic, the development of robust workflows operating on data and their reproducible execution is a central component in any scientific investigation.
To facilitate this process for users of \signac, the \sflow\ package provides users with a flexible set of tools to implement workflows operating on \signac\ data spaces (see \cref{sec:signac-flow}). 

\begin{figure*}[t]
    \includegraphics[width=\linewidth]{{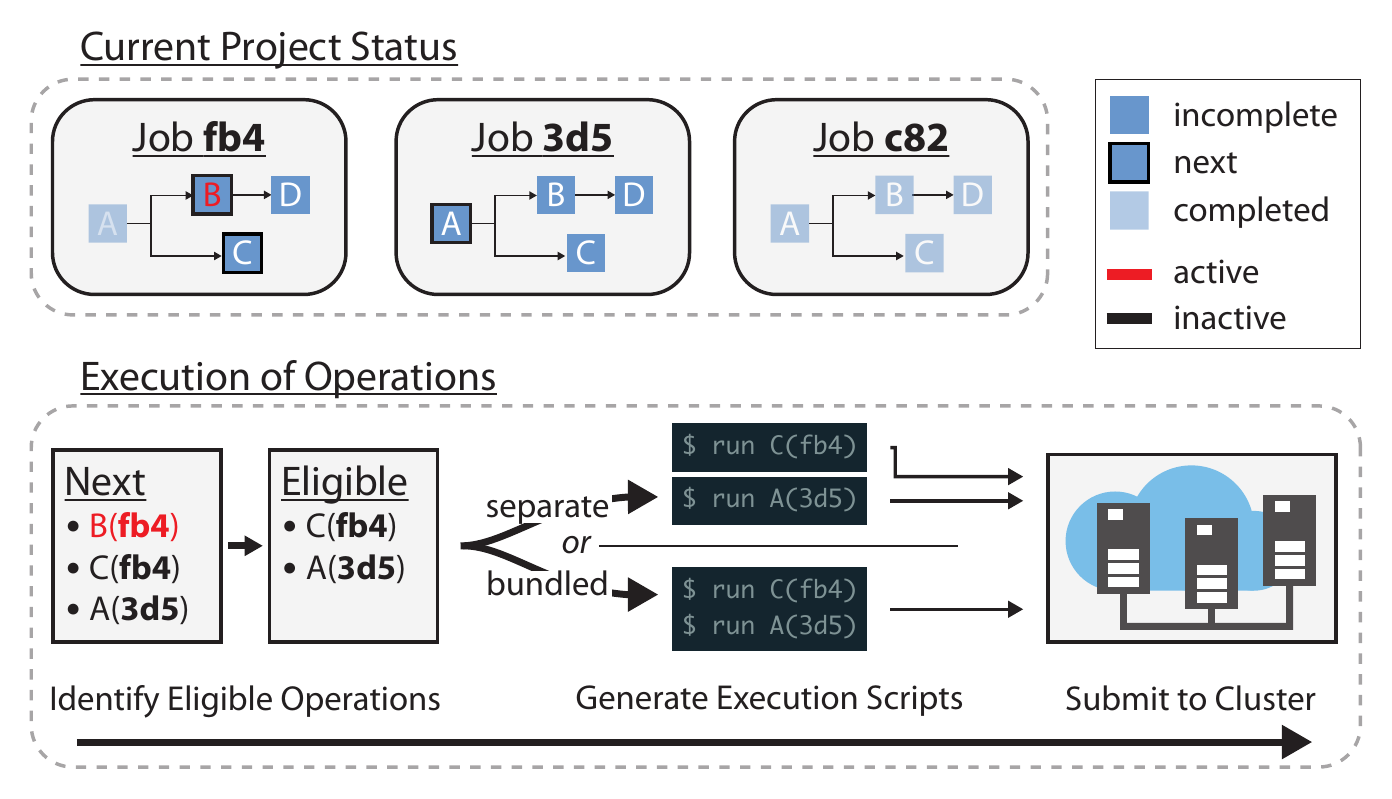}}
    \caption{
    In order to track and execute workflows on a \signac\ workspace, \sflow\ \fps\ track the status of each job (top).
    This status tracking includes information about which operations have been completed for a given job, which operations are next in line to run, and which operations are incomplete but are not ready to run due to unfulfilled dependencies upstream in the workflow.
    The progression of each job through the workflow is always known to the \fp, as is whether a particular job-operation pair is \emph{active}, \textit{i.e.}, is currently executed on a \glsdesc{hpc} cluster or is queued for execution.
    This information is used to determine which job-operation pairs are \emph{eligible} for submission to the cluster scheduler; pairs that are already queued or active are not resubmitted (bottom).
    For maximal flexibility, the execution of job-operations may be \emph{bundled} prior to submission, enabling, \textit{e.g.}, the execution of large numbers of compute-light operations on a single node in serial or parallel.
    }
    \label{fig:signac-flow}
\end{figure*}

\section{Implementation}

\subsection{Software Architecture}

The core \signac\ data management application, as well as the rest of the \signac\ framework, are implemented in Python and tested for versions 2.7.x and 3.x.
They are designed to be used in \glsfirst{hpc} environments, and hard requirements besides the Python interpreter are avoided.
We employ continuous integrated testing to ensure high interoperability between all main applications.
Documentation is generated via the Sphinx documentation tool~\cite{thepocooteam2016} and made available online\footnote{signac.readthedocs.io and signac-flow.readthedocs.io}.

Although the primary interface is Python-centric, most core \signac\ functionality is available through a \gls{cli} to simplify the integration of workflows that are not Python-oriented.
Metadata is encoded in the open standard \gls{json} format, which is largely human-readable and can be easily parsed in most programming languages.
Relying on a simple, open format ensures that the data remains accessible even without \signac.
Furthermore, the \gls{json} format is internally used by many \gls{nosql} \glsfirst{dbms}, allowing an effortless integration of \signac\ with these systems.

\subsection{Software Components}
\label{sec:components}
The main data management functions of the \signac\ framework are implemented as part of the core \signac\ application.
This application is designed with modularity in mind, enabling its extensibility \emph{via} the implementation of additional components of the \signac\ framework.
This layered structure minimizes the interdependence of higher-level components, making the system more robust against architectural changes~\cite{Martin2012}.
Besides the main application, we have implemented various other (partially not yet published) tools to augment the \signac\ ecosystem such as the \sdashboard, a web application to search and visualize \signac\ data spaces in the browser.

In this section, we first describe the three primary functions of the core application: data storage and searching, which simplifies the maintenance and access of complex and heterogeneous data spaces; indexing, which enables efficient advanced post-processing and analysis routines; and database integration, which allows the export of indexes and data to external databases.
We then demonstrate this framework's extensibility in \cref{sec:signac-flow}, where we discuss the \sflow\ application that we have developed for the management of workflows utilizing \signac's data management capabilities.

\subsubsection{Project Data Management}
\label{sec:data-interface}

The data management component is the central component of the \signac\ model.
The framework supports all typical data management related processes, including data curation, data manipulation, and analysis, by providing a consistent and homogeneous interface for data access and storage within the workspace.
The workspace itself is project agnostic, \emph{i.e.}, the particular workspace associated with a project may be swapped in and out at any time, and workspaces can be divided and merged as depicted in \cref{fig:workspaces-indexing}.

The main challenge of reliable long-term storage of data is to ensure the proper association of data and metadata.
To surmount this obstacle, the \signac\ application calculates a short numeric hash value from the full parameter metadata to generate a unique address, the \textit{signac id}, which is a concise representation of the full state point.
The signac id serves as the primary index and constitutes the basis for the file system path within the workspace where associated data is stored.
A \gls{json}-encoded copy of the parameter metadata is saved within these paths, which ensures that this association can be trivially identified.
The use of a standard format such as \gls{json} ensures that access to the data is not dependent on \signac.

This methodology bypasses numerous issues common to file system-based workflows.
As the output of a hash function, the signac id is both short and non-ambiguous, making it a unique, reliable, and indexable address of the data in all contexts.
The signac id can also encode effectively arbitrary complexity, circumventing file naming limitations inherent to most file systems while maintaining great flexibility.

\subsubsection{Indexing and Database Integration}
\label{sec:indexing}
The internal index that \signac\ generates to support its main functions is exposed to the user \emph{on demand}.
This can be used to simplify the mapping between different, possibly heterogeneous storage devices, such as a file system and a database system.
For example, we could use \signac\ to generate files on the file system and execute post-processing routines on the data, and then export the data index into a database that is accessible to a wider group of data subscribers.

To facilitate integration, the current implementation supports export routines for the MongoDB \gls{nosql} database, but in principle any database system that provides a Python driver could be integrated in the future.
We chose to initially support MongoDB because its internal data structure is already based on the \gls{json} format and because we consider the semi-structured \gls{nosql} approach more flexible and intuitive to researchers, who are used to dynamic schemas rather than the more rigidly defined table schema used in relational \gls{dbms}.
Using MongoDB also enables users to leverage tools built for the MongoDB ecosystem for data inspection and manipulation, \textit{e.g.}, \package{Studio 3T}~\cite{studio3t}.

The indexes in \signac\ are generated by one or more \emph{crawlers}, which for our purposes are defined as any functions that generate a series of \gls{json} documents.
In general, the index needs to contain the metadata associated with the data and all information required to allow access to the data.
In the specific case of a file system index, this is the metadata and information about file locations and formats.
The system is designed for simple customization, \textit{e.g.}, for the extraction of additional metadata from the data (deep indexing).
The \signac\ application provides templates for crawlers specialized to crawl file systems and generate indexes.

The data processing and index creation steps are intentionally decoupled in \signac, allowing easy indexing of pre-existing data.
This approach is enormously powerful in providing a single homogeneous data interface for new and existing data, particularly because crawlers can be used to index data spaces not generated by \signac.
These indexes can be used to make data accessible to individual researchers within and across organizations, whether or not \signac\ was used for their curation.

\subsection{Implementation of workflows with \sflow}
\label{sec:signac-flow}

Although \signac\ is designed to be workflow agnostic, it is very important for computational scientists to maintain a well-defined workflow that interacts in predictable ways with data.
To ease the development of computational workflows using \signac, we developed the \sflow\ package, which offers users the ability to design complex workflows around \signac\ managed data spaces (illustrated in \cref{fig:signac-flow}).
There are three critical elements of \sflow: \emph{jobs}, each of which represents the data associated with a single parameter combination; \emph{operations}, which are sets of procedures acting on jobs; and \emph{\fps}, which are collections of operations encapsulating a complete workflow associated with a \signac\ data space.
Note that \fps, which correspond to a single workflow, are distinct from \signac\ projects, which correspond to a particular data space.
The \sflow\ package supports multiple \fps\ acting on a single \signac\ project to allow the implementation of multiple distinct workflows on the same data space.
An example of where this might be useful would be to create separate \fps\ to perform coarse-grained and atomistic molecular dynamics simulations of the same system to extract different sets of information.

To convert our original ideal gas workflow into a \sflow\ \fp, we could define an \texttt{IdealGasEquationOfState} \fp\ with a single operation responsible for calculating the volume from the parameters.
If we desired, we could then easily define additional operations for, \emph{e.g.}, the computation of the free energy of the gas.
For more complex workflows, the sequence is controlled by a series of pre- and post-conditions for each operation that determine the next set of operations that should be executed.
The \fp\ is entirely self-contained, relying on \signac\ to store and manage the generated data.

The \sflow\ package is also designed to facilitate working with compute clusters.
For this purpose, we define a \emph{job-operation} as an atomic task consisting of a \fp\ operation acting on a specific job. 
The \fp\ interface enables the packaging of sets of job-operations into cluster jobs by automatically generating the requisite job scripts; each cluster job can consist of an arbitrary number of job-operations running either in serial or in parallel.
At the time of writing, \fps\ support submission to both Slurm and Torque PBS clusters, generating job scripts on-the-fly after detecting the types of job schedulers present on a given cluster.
The \sflow\ package allows users to configure their default submission behavior, both globally and on the level of a single \fp.
In addition, users working in cluster environments with specific requirements, such as submitting only to a specific partition, can encapsulate this information into specific Python modules that \sflow\ can be configured to recognize, making it easier for users to share common configuration information.
By providing simple and transparent APIs for cluster submission, \sflow\ enables users to streamline the large-scale execution of data space operations in cluster environments.

\section{Practicality and Scalability}
\label{sec:comparison}

To assess the practicality and scalability of our implementation, specifically with respect to existing comparable solutions, we evaluated the following key metrics:
\begin{enumerate}
    \item Efficiency of setting up a new workflow for an existing tool set.
    \item Time needed to determine the data space size.
    \item Time needed to iterate through the data space.
    \item Time needed to search and select data sets.
\end{enumerate}
Since the first item is difficult to \emph{quantify}, we instead attempt to demonstrate how easily \emph{any} scriptable tool operating on input and output files may be integrated into a \signac{}- and \sflow-based workflow by means of the examples laid out in \cref{sec:examples}.
The remainder of this section is dedicated to a more in-depth comparison of \signac\ with alternative solutions, including benchmarks for the last three items in direct comparison with datreant, which we identified as the most comparable tool in both scope and approach.

\subsection{Comparable Solutions}
\label{sec:survey}

The \signac\ philosophy entails leaving the development of data schemas and workflows largely up to the user, removing the need for specialized input scripts and output parsers.
In this way, \signac\ substantially differs from domain-specific tools such as AiiDA~\cite{Pizzi2016} or pylada-light~\cite{pylada-light}, which impose strict data and workflow restrictions.
We believe that this relaxed structure decreases the barrier for integrating new tools and developing new workflows; however, we also recognize that this less standardized approach increases the chance of user error during the implementation and execution of workflows.

In the realm of workflow management, the FireWorks open-source tool~\cite{FireWorks} stands out as a particularly mature and feature-rich option.
Its feature set largely overlaps with the one provided by \sflow, and in addition it offers more advanced job management and monitoring capabilities.
These additional features are supported by a MongoDB database on the back-end.
In contrast, \sflow\ relies purely on \signac\ to store all runtime and scheduling related metadata.

Integrating FireWorks and \signac\ simply involves using \signac\ to manage the data space while specifying and executing workflows through the FireWorks interface, similar to how \sflow\ is currently integrated with \signac.
Yet, there is a caveat: FireWorks' data storage layout is strongly coupled to its execution model, to the extent that Fireworks' documentation explicitly discourages users from manually controlling data storage locations\footnote{https://materialsproject.github.io/fireworks/controlworker.html}.
The tools operate on two different philosophies when it comes to storage layout management, which poses a barrier for integration.

The Sumatra tool~\cite{sumatra} allows users to keep a detailed ``automated electronic lab notebook''\footnote{http://neuralensemble.org/sumatra/} of operations executed on a specific data space.
It is not a job manager in the sense of FireWorks or \sflow, but primarily focuses on ensuring that computational research is reproducible.
We found it to integrate very well with \signac\ and \sflow\ operations, enabling users to keep better track of which operations have been executed, a feature which \sflow\ currently lacks.

The software we found to be most similar to \signac\ in core scope and functionality is datreant.core~\cite{datreant}, which enables users to associate specific directories with searchable metadata.
Just like \signac, datreant.core is largely domain agnostic, does not require a central server, and performs distributed data management directly on the file system in distinct directories that are associated with searchable metadata.

However, there are also some key differences.
First, datreant.core is even more agnostic than \signac\ with respect to the general workflow, \textit{e.g.}, there is no need to confine data within a single project entity.
Instead, multiple directories may be dynamically organized in \emph{bundles}, which loosely correspond to a \signac\ workspace but need not share a common root directory.
These bundles can be searched and grouped by metadata, just like \signac\ jobs.
Furthermore, datreant.core has no concept of a unique identifier like the \emph{signac id}, so the user is still required to choose a directory name for each data set.
While this methodology might provide more flexibility in defining a general storage layout and make it easier to combine different data spaces, we contend that it would make it harder for novices to overcome the habit of encoding metadata in file paths, reducing the homogeneity and flexibility of the overall data space.
Finally, datreant.core employs file locking mechanisms to ensure that metadata may be safely manipulated in parallel from multiple processes.
While that might be advantageous under some circumstances, in practice file locks do not work reliably on the network file systems commonly employed in \gls{hpc} environments, rendering this feature a liability in cases where it would be most needed.
For this reason, the \signac\ implementation avoids any reliance on file locks.
Overall, we have found datreant to be the most comparable existing solution for the core problems \signac\ aims to solve.

\subsection{Benchmarks}
\label{sec:benchmarks}

\begin{figure}
    \centering
    \includegraphics[width=\linewidth]{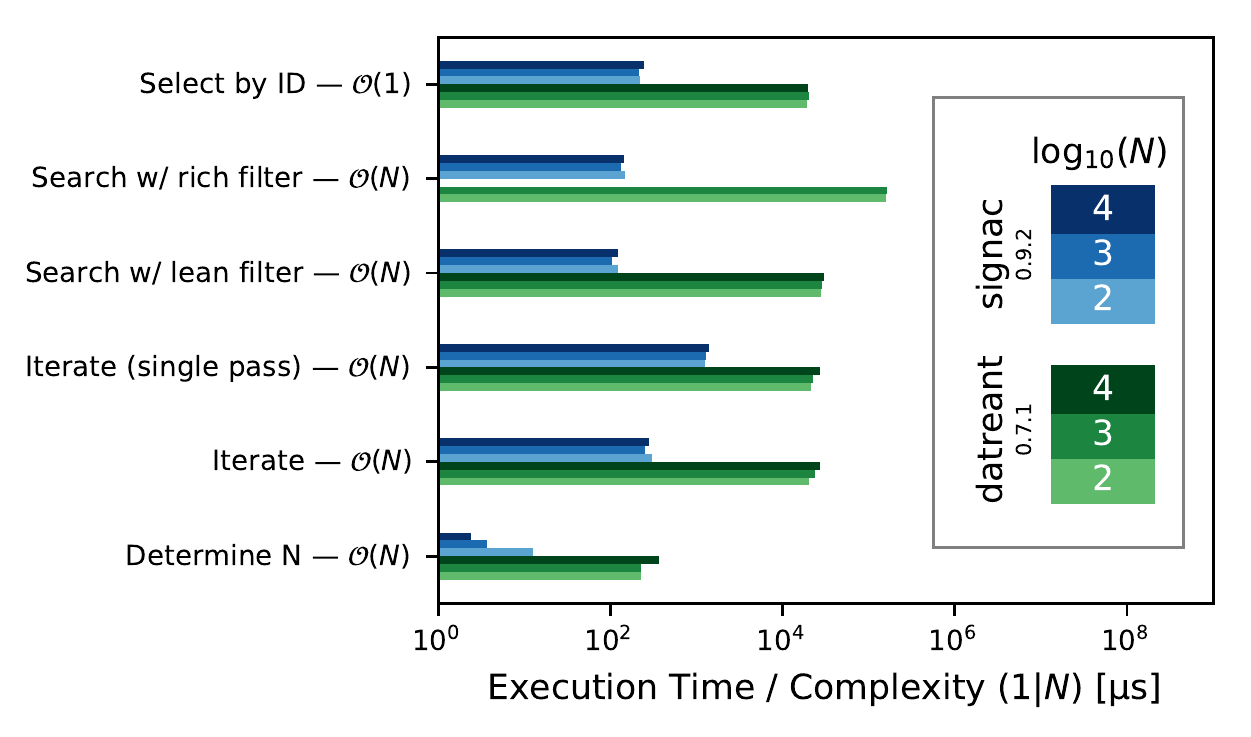}
    \caption{
        We measured the time required for the execution of a set of data space operations as a function of the number of directories $N$ with \signac\ and datreant.
        All tests were executed with Python 3.6 on a network file system;
        reported values are the minimum of 3 independent test sessions, where each one is averaged over 10 runs within one session, except for the 4th, which was run only once per session;
        the 2nd test category was aborted for datreant at $N=10^4$ due to very long execution time.
        All values are normalized by the expected complexity, \textit{i.e.}, they must be multiplied with the respective order to obtain absolute values for a specific data space size.
        }
    \label{fig:benchmark}
\end{figure}

Since datreant.core most closely  corresponds to \signac's scope and approach for data management, we used it as a quantitative benchmark for the performance and scalability of our implementation.
Concretely, we measured the time each tool required
\begin{enumerate}
    \item to select a single data set by known id,
    \item to search and select with a rich filter (many keys),
    \item to search and select with a lean filter (one key),
    \item for the first iteration through the meta data space within one session,
    \item for multiple iterations within one session,
    \item to determine the data space size $N$.
\end{enumerate}
We then used this data to estimate the time complexity of each operation with respect to $N$.
In \signac, we expect all but the first category to run with a time complexity of $\mathcal{O}(N)$, since the largest bottleneck is likely to be the initial parsing of all metadata within one session.
A \emph{selection by known id} should be constant time $\mathcal{O}(1)$.
The results of these measurements are plotted in \cref{fig:benchmark}.

We are able to show that within our test environment\footnote{A workstation with 20 Intel(R) Xeon(R) CPU E5-2680 v2 @ 2.80GHz cores running Gentoo Linux (4.9.34).} a data space of $N=1,000$ directories and approximately \SI{1}{\kilo\byte} of metadata per directory\footnote{That corresponds roughly to 10 keys of one character associated with 100 character long values.}, all operations, even those that scale linearly with the data space, are executed nearly instantaneously on a human time scale.
For example, the first iteration through the complete metadata space within one session requires on the order of \SI{1}{\milli\second} per directory.
It is important to point out that none of these operations are in any way affected by the number or size of data files within those directories since they only interact with the \gls{json} metadata files.

While both \signac\ and datreant show very similar scaling behavior, we can clearly show that \signac\ is at least one order of magnitude faster than datreant in all tested categories despite implementing very similar concepts.
The maximum practical data space size -- at which users perceive the \gls{srt} acceptable for complex tasks  -- is therefore much larger. 

Comparing our time measurements on a network file system ($\sim$\SI{1}{\milli\second} per directory for start-up) with the guidelines laid out by Doherty and Sorenson~\cite{Doherty2015}, operating on data spaces with up to 300 directories would be perceived as instantaneous ($<$\SI{300}{\milli\second}), 1,000 directories as immediate ($<$\SI{1}{\second}) and up to 5,000 directories as transient ($<$\SI{5}{\second}).
Larger data spaces with up to 300,000 directories may still be acceptable, but will require multitasking and/or additional feedback on the progress to not break the user flow.

In summary, while the only hard cap on the data space size is the file system and main memory storage capacity, interactive work may be significantly impaired by prolonged session start-up times for data spaces with more than 300,000 directories.
In this case users would be advised to aggregate the working set of data prior to interactive work.
We consider data spaces with up to 10,000 directories very practicable for interactive work even on network file systems.
All the code to generate these benchmarks is open source and available online\footnote{https://bitbucket.org/glotzer/signac-benchmarks}.

\section{Examples}
\label{sec:examples}

\lstset{
    basicstyle=\footnotesize\ttfamily,
    language=python,
    numbers=left,
    numbersep=0.5em,
    tabsize=2,
    showstringspaces=false
    }

In this section we introduce two representative conceptual examples that demonstrate how to incorporate \signac\ into computational workflows.
The first one is in reference to the case presented in \cref{sec:introduction}, the evaluation of the equation of state of an ideal gas.
The second is a molecular dynamics study of the Lennard-Jones potential, which is slightly more involved, but also more realistic.

For brevity, some commands are omitted or shortened; however, fully functional examples, including additional demonstrations for DFT and GROMACS, can be found in the supplementary material or online\footnote{https://bitbucket.org/glotzer/signac-examples}.
All Python examples are tested for Python version 3.5.

\subsection{Ideal Gas Example}

This is a minimal demonstration for carrying out the example described in the introduction.
We intend to calculate and store the volume $V$ of an ideal gas within the three-dimensional parameter space spanned by $p$, $N$, and $kT$.

We start by creating an empty directory for our project and initializing the signac project:
\begin{lstlisting}[language=bash,numbers=none]
$ mkdir idg_eos
$ cd idg_eos
$ signac init IdealGasEOS
\end{lstlisting}
The project initialization creates a small configuration file within the current directory to mark it as the project's root directory.

\subsubsection{Minimal Ideal Gas Example}

For our most basic demonstration, we implement a Python script to calculate and store the volume in \signac's built-in \gls{json} storage for each state point of interest:
\begin{lstlisting}
import signac

project = signac.get_project()

for p in 0.1, 1.0, 10.0:
  sp = {"p": p, "N": 1000, "kT": 1.0}
  job = project.open_job(sp)
  V = job.sp.N * job.sp.kT / job.sp.p
  job.document["V"] = V
\end{lstlisting}
First, we import the \lstinline{signac} Python package (l.\,1).
Then we obtain a handle on the project (l.\,3), which is the interface for accessing and manipulating the project's data space.
To calculate the phase diagram --- here as a function of pressure --- we simply iterate over $p$ (l.\,5) and construct the full state point \lstinline{sp} associated with each data point (l.\,6).

This state point is passed into the \lstinline{project.open_job()} function, which returns a \emph{job handle} that represents this specific data point (l.\,7).
The volume is calculated from the state point variables associated with the job, which we access \textit{via} the \lstinline{job.sp} property (l.\,8).
Being a single number, the volume naturally lends itself to being stored in a very lightweight format.
Here, we leverage the \lstinline{job.document} property of \signac\ jobs, which provides a lightweight, persistent, and immediately searchable \gls{json} storage option associated with each \signac\ job  (l.\,9).
However, we could store the data just as well in a file with a format of our choosing, as will be shown in the next example.

Once the data space is initialized, we can immediately start searching it.
For example, to find all state points, where \emph{$p$ is greater than 1.0}, we would execute:
\begin{lstlisting}[numbers=none]
jobs = project.find_jobs({"p.$gt": 1.0})
\end{lstlisting}
The \lstinline{jobs} variable is the result cursor that we can use to iterate over all jobs that match the given criterion.
We can execute the same kind of queries directly on the command line:
\begin{lstlisting}[language=bash,numbers=none]
$ signac find p.\$gt 1.0
\end{lstlisting}
In this case the ids of all matching jobs will be output for further processing.
The query language supports a variety of operators, including but not limited to arithmetic and logical operators, and  represents a subset of the \mongodb\ query language, making it easy to transition between the two systems.
More details can be found in the online documentation.

\subsubsection{Ideal Gas with a Bash Terminal Script}
\label{sec:example-idg-bash}
In many cases parts of our workflow will rely on precompiled programs or other scripts that can be interfaced on the command line, but not directly through Python.
For example, we might have a program called \lstinline{idg}, that accepts parameters $N$, $kT$, and $p$ as the first through third argument and outputs the resulting volume $V$, \textit{e.g.}:
\begin{lstlisting}[numbers=none]
$ idg 1000 2.0 1.0
2000.0
\end{lstlisting}

The \signac\ application provides a \glsfirst{cli} to simplify the integration of such tools.
The following example script replicates the first example, but in bash instead of Python and the volume is stored in a file called \lstinline{V.txt} instead of the job document.
\begin{lstlisting}[style=mystyle]
#!/bin/bash
N=1000
kT=1.0
for p in 0.1 1.0 10.0; do
  SP={\"N\": $N, \"kT\": $kT, \"p\": $p}"
  WS=@\backtick@signac job -wc "$SP"@\backtick@
  ./idg $N $kT $p > $WS/V.txt
done
\end{lstlisting}
After storing parameters as constants at the beginning of the script (l.\,2-3), we again iterate over the variable of interest (l.\,4) and construct the full state point \lstinline{SP} in \gls{json} formatting\footnote{The \gls{json} format expects all keys to be enclosed in double quotes, which need to be escaped within the bash script.
We recommend using here-docs for larger state point definitions.} (l.\,5).
We then provide the state point as argument to the \lstinline{signac job -wc} command, which creates the corresponding job and returns the full workspace path \lstinline{WS} (l.\,6).
Finally, we execute the \lstinline{idg} program and pipe its output into the \lstinline{V.txt} file within the job's workspace (l.\,7).
This approach reliably couples the job's data and the parameters used to generate them.

An alternative approach for the incorporation of command line tools is the construction of the required bash commands within a Python script:
\begin{lstlisting}
import signac
from subprocess import run

IDG = "./idg {job.sp.N} {job.sp.kT} {job.sp.p}"\
      ">{job.ws}/V.txt"

project = signac.get_project()

for p in 0.1, 1.0, 10.0:
  sp = {"N": 1000, "kT": 1.0, "p": p}
  job = project.open_job(sp)
  job.init()
  if not job.isfile("V.txt"):
    run(IDG.format(job=job), shell=True)
\end{lstlisting}
This approach can be more flexible, especially in cases where users are already familiar with Python.
The crucial point is that input parameters and location of the output data are always automatically and unambiguously linked.

\subsection{Molecular Dynamics with HOOMD-blue}

Similar to the first example, we again calculate the equation of state of a gas, this time using molecular dynamics with a Lennard-Jones potential.
This means that instead of merely evaluating a single analytic function, we need to set up initial and boundary conditions of the simulated system, load the interaction potential, define the simulation protocol, and possibly store significant amounts of output data.

\subsubsection{Basic example}

For this example we will use the \hoomd ~\cite{Anderson2013a,Anderson2008,Glaser2014b} particle simulation toolkit which provides a native Python interface.
This means we can interface with the \signac\ project  directly within the input script.
If there was no Python interface, we would follow the approach shown in the previous (CLI) example (\cref{sec:example-idg-bash}).
\begin{lstlisting}
import signac
import hoomd
import hoomd.md

def setup_and_simulate(job):
  # [...] Setup initial conditions
  hoomd.md.integrate.langevin(
    kT=job.sp.kT, seed=job.sp.seed, ...)
  hoomd.dump.gsd(
    filename="trajectory.gsd", period=2e3, ...)
  hoomd.run(steps=1e4)

project = signac.get_project()

for kT in 0.1, 1.0, 2.0:
  sp = {"kT": kT, "seed": 42}
  with project.open_job(sp) as job:
    setup_and_simulate(job)
\end{lstlisting}
We start by importing all required packages (l.\,1-3) and continue by defining a function for the execution of our simulation as function of the job (l.\,5).
We skip HOOMD-specific commands needed for the setup of the simulation, but lines 7 and 8 show how we use the \lstinline{job.sp} interface to directly set the simulation parameters.

The iteration over the data space (l.\,15) and the definition of the full state point (l.\,16) are analogous to the previous examples.
Instead of wrapping all input and output filenames wherever they appear (such as in line 10), we use \signac's built-in context manager to change into the job's workspace for all commands that are within the scope of the \lstinline{with} clause (l.\,17).
That means \signac\ will change into the correct directory for the duration of the execution of the \lstinline{setup_and_simulate()} function and return to the previous directory after completion.

In this example, the data space operations that we execute are still very simple: simulations are executed sequentially by iterating over the variable of interest, $kT$.
However, for more complex workflows, especially those involving more compute-intensive operations, it is advantageous to break things up into smaller steps that can be executed in parallel and possibly be submitted to an \gls{hpc} cluster.
One possible approach for doing so is shown in the next example, utilizing the the previously introduced \sflow\ application (see \cref{sec:signac-flow}).

\subsection{Workflow management with \sflow}

While users are encouraged to integrate the \signac\ data management application into existing workflows or develop new ones that fit their specific applications, here we demonstrate the use of the \sflow\ application for the rapid development of workflows for users that are so inclined.
The application is quite general, and is simply designed around the sequential or parallel execution of operations in well-defined order.
Splitting the overall workflow into such self-contained operations increases flexibility and reproducibility and is especially beneficial for larger studies.

We demonstrate the concept by adapting the previous example.
First, we move the initialization logic into a separate script to \emph{initialize} the data space prior to executing any data space operations:
\begin{lstlisting}
# init.py
import signac

project = signac.init_project("LJ-EOS")

for kT in (0.1, 1.0, 2.0):
  sp = {
    "kT": kT, "seed": 42,
    "epsilon": 1.0, "sigma": 1.0,
    "r_cut": 3.0}
  project.open_job(sp).init()
\end{lstlisting}
This initializes the complete data space with the essential parameters required for the execution of our molecular dynamics simulations.

Second, we split the \lstinline{setup_and_simulate()} step into \lstinline{setup()} and \lstinline{simulate()}.
These two operations are defined within an \lstinline{operations.py} module:
\begin{lstlisting}
# operations.py
import hoomd
import hoomd.md

def setup(job):
  """Setup the initial conditions"""
  hoomd.init.create_lattice(
    unitcell=hoomd.lattice.sc(a=1.0), n=16)
  hoomd.dump.gsd(
    filename=job.fn("init.gsd"), ...)

def simulate(job):
  """Execute MD simulation"""
  with job:
    hoomd.init.read_gsd("init.gsd")
    # [...]
    lj = hoomd.md.pair.lj(r_cut=job.sp.r_cut, ...)
    lj.pair_coeff.set(
      "A", "A",
      epsilon=job.sp.epsilon,
      sigma=job.sp.sigma)
    hoomd.md.integrate.langevin(
        kT=job.sp.kT, seed=job.sp.seed, ...)
    hoomd.dump.gsd(
        "trajectory.gsd", period=2e3, ...)
    hoomd.run(tsteps=1e6)
    job.document["step"] = hoomd.get_step()

if __name__ == "__main__":
  import flow
  flow.run()
\end{lstlisting}
The last three lines (l.\,29-31) leverage \sflow's function to equip this module with a command line interface that allows us to execute all operations directly from the command line:
\begin{lstlisting}[language=bash,numbers=none]
$ python operations.py setup
$ python operations.py simulate
\end{lstlisting}

To further automate the execution of operations and their submission to an \gls{hpc} cluster, we can implement a workflow as part of a \fp\ as described in \cref{sec:signac-flow}.
The workflow is defined by adding operations to the \fp\ class with \lstinline{add_operation()} during its construction.
Each operation can be associated with pre- and post-conditions to determine their order of execution.
An operation is eligible to be executed when all pre-conditions are met and at least one of the post-conditions is not met.
The execution conditions associated with each operation are implemented as methods, which are then passed as arguments to the \lstinline{pre} and \lstinline{post} parameters of the \lstinline{add_operation()} method.

For this example, we would want to execute the \lstinline{setup} operation first, and then assuming that was successful, the \lstinline{simulate} operation.
A simple condition for successful setup would therefore be the existence of the \lstinline{init.gsd} file, which contains the system's initial configuration, so we set that as the \texttt{post} condition \emph{via} the \texttt{initialized} function.
We also keep track of the \emph{simulation progress} by storing the current simulation time step within the persistent \gls{json} storage associated with the job (\lstinline{job.document}).

\begin{lstlisting}
# project.py
import flow

class MyProject(flow.FlowProject):

  def initialized(self, job):
    return job.isfile("init.gsd")
    
  def simulated(self, job):
    return job.document["step"] >= 1e6
  
  def __init__(self, *args, **kwargs):
    super().__init__(*args, **kwargs)
    
    # Add the "setup" operation
    self.add_operation(
      name="setup", 
      cmd="python operations.py setup {job._id}",
      post=[self.initialized])
      
    # Add the "simulate" operation
    self.add_operation(
      name="simulate",
      cmd=\
        "python operations.py simulate {job._id}",
      pre=[self.initialized],
      post=[self.simulated])
      
if __name__ == "__main__":
  MyProject.main()
\end{lstlisting}
In addition to clearly defining the status of each individual operation, this \fp\ implementation also describes all valid sequences of operations.
For example, because the \lstinline{setup} operation has no pre-conditions, it is the only operation eligible for execution immediately after data space initialization.
Since the \fp\ encapsulates this logic, we can trivially execute our workflow by leveraging the \fp's \emph{run} capabilities, which take the simple run functionality from our \lstinline{operations.py} script one step further.
Rather than specifying operations to run, we can now simply execute \lstinline{$ python project.py run}, which will automatically run the next eligible operation for each job.
To submit these job-operations to a job scheduler on a \gls{hpc} cluster, we could instead use \sflow's submission tool by typing \lstinline{$ python project.py submit}.
The objective of dividing the implementation of operations and the definition of workflows as part of the \fp\ is to avoid the conflation of responsibilities and to ensure a very clear path for the integration of operations that are not Python-based.

Complete versions of the examples presented here, as well as some additional ones, can be found as part of the supplementary material and online\footnote{https://bitbucket.org/glotzer/signac-examples}.

\section{Conclusions}

The development of \signac\ is motivated by the increased need for the management of heterogeneous and complex data spaces in computational materials science, specifically in work requiring \gls{hpc} resources.
Researchers in computational fields are frequently required to manage such data spaces and account for the various issues associated with this task.
The \signac\ framework provides non-intrusive solutions to many data management and workflow challenges in environments scaling from desktops to \gls{hpc} clusters.
The simple file-centric data model and the use of standard file formats such as \gls{json} ensure easy access and portability of both the data and the associated workflows.
This portability is particularly critical for sustainable long-term storage, since it allows the use of \signac\ without tying users to future use of the platform or specific file formats in order to be able to access the data. 
The indexing functionality eases the transition from data acquisition to curation and analysis, and the simplicity of export to databases allows the integration of existing \gls{dbms} into \gls{hpc} workflows.
These functions allow \signac\ to combine the advanced metadata handling capabilities of modern \gls{dbms} while also providing the performance of file system-based solutions.
By providing a lightweight, high-performance solution to common data management and workflow challenges in \gls{hpc}, the \signac\ framework frees researchers from solving these problems themselves and enables more effective and efficient scientific research.

\section*{Acknowledgments}

\small{
We would like to thank all contributors to the development of the framework's components, J.A. Anderson, M.E. Irrgang and P.F. Damasceno for fruitful discussion, feedback and support,
and B. Swerdlow for his contributions and feedback and coming up with the name.
Finally, we would like to thank all early adopters that provided feedback and thus helped in guiding and improving the development process.
Development and deployment supported by MICCoM, as part of the Computational Materials Sciences Program funded by the U.S. Department of Energy, Office of Science, Basic Energy Sciences, Materials Sciences and Engineering Division, under Subcontract No.~6F-30844.
Project conceptualization and implementation supported by the National Science Foundation, Award~\#~DMR~1409620.
}

\section*{References}

\bibliographystyle{elsarticle-num-names}
\bibliography{signac-paper}
\end{document}